\documentclass[doublespacing]{elsart}
\usepackage{natbib}
 \usepackage{graphicx}
\usepackage{epstopdf}

\begin{document}
\runauthor{Coward}
\begin{frontmatter}
\title{Open issues with the gamma-ray burst redshift distribution}
\author{David Coward}

\address{School of Physics, University of Western
Australia, M013, Crawley WA 6009, Australia, Email: coward@physics.uwa.edu.au}

\begin{abstract}
Cosmological gamma ray bursts (GRBs) are the brightest explosions in the Universe. Satellite detectors, such as Beppo-SAX, HETE2 and more recently {\it Swift}, have provided a wealth of data, including the localization and redshifts of subsets of GRBs. The redshift distribution has been utilized in several studies in attempts to constrain the evolving star formation rate and to probe GRB rate evolution in the high-redshift Universe. These studies find that the GRB luminosity function and/or the rate density evolve with redshift. We present a short review of the problems of constraining GRB rate evolution in the context of the complex mix of biases inherent in the redshift measurements. To disentangle GRB rate evolution from the biases prevalent in the redshift distribution will require accounting for the incompleteness of the observed redshift sample. We highlight the importance of formulating a `complete GRB selection function' to account for the main sources of bias.
\end{abstract}


\end{frontmatter}

\section{Introduction}
Cosmological gamma ray bursts (GRBs) are the brightest explosions in the Universe. They show a bi-modal
distribution in durations of short bursts of around 0.3 s length
and long bursts of around 30 s \citep{kou93}. Their fluence
(time-integrated flux) and time-variability implies that their
inner engines must be compact and highly energetic. 
The association of some long-duration or `classical' GRBs with supernovae implies that they are associated with the core collapse of massive stars. The few short bursts that have been well localized show no supernova association and are associated with both early and late type galaxies ---see Berger (2006). Short bursts most likely have a different evolutionary channel compared to long GRBs, one possible scenario being  the merger of neutron star--neutron star or neutron star--black hole binaries. In the following discussion, we focus on observations of long GRBs, hereafter just GRBs, from the suite of satellite detectors, and  the redshifts obtained from a subset of localized GRBs.

Because of their extreme luminosities in $\gamma$-rays, GRBs are a promising new probe of the high-$z$ Universe. They provide the opportunity to study their local interstellar environment, host galaxies and the intergalactic medium. The association of GRBs with short-lived massive stars implies that they may track the evolving star formation rate (SFR). These exciting possibilities provide strong motivation to use the measured GRB redshift distribution as a tool to constrain the SFR and GRB rate evolution.

To constrain GRB rate evolution requires an accurate model for the GRB luminosity function (LF), the selection function of the GRB satellite detectors and the influence of using different ground-based telescopes and techniques with varying quality to measure redshifts from the GRB afterglow. Complicating the analysis further is the assumption that GRBs represent one unified population with a distribution of luminosities encompassed by a single luminosity function. Determining the effects of these biases on the redshift distribution is fundamental to constraining the astrophysical GRB parameters. Here, we provide an overview of recent work on constraining GRB rate evolution and the SFR using the measured GRB redshift distribution. We describe from first principles the flux-limited GRB selection function and follow this with a discussion on the `complete' GRB selection function, which encompasses all selection effects but is presently poorly understood.

\subsection{The evolving star formation rate}
Several observation-based estimates of the evolving SFR have been
made in recent years, mostly using the
observed rest-frame UV luminosity density of the galaxy
population. However, there has not been a consensus amongst
workers on the form of SFR evolution because of systematic
uncertainties, such as the amount of dust extinction, and because
of the difficulties of the observations. Early studies of the UV
and IR luminosity densities of star-forming galaxies
\citep{Lill96,madau96,conn97,M1998} 
suggested that the SFR
increases steeply from $z= 0$ to $z\approx 1-2$, and then declines
at higher $z$. However, 
\citet{steid99} 
favour a near-constant SFR density
at $z\approx1 - 4$, and 
%\citet{Lan2002} 
model a monotonically
increasing rate out to $z=10$. Contrary to some of these
results, 
\citet{sull2001} 
find a more gentle increase in the rate.

\citet{hb06} have constrained the SFR by fitting to ultraviolet and far-infrared data obtained up to 2006. When these data are taken together, they show that it is possible to constrain the SFR out to $z \approx 6$, with especially tight constraints for
$z < 1$. Figure 1, taken from their work, shows that the SFR is well constrained from $z=0-1$ but is not well defined at $z>2$, as shown by the larger error bars and smaller density of data. These data indicate that the SFR may be either roughly constant or declining at $z>5$.

Because of the large uncertainty in the SFR at $z > 1$, simple parametrized models with 3 free parameters have been used as fitting functions. \citet{M2000} and \citet{P2001} provide observation-based
parametrized SFR density models for three scenarios. The first
matches most measured UV continuum and H$\alpha$ luminosity
densities and includes an upward correction for dust extinction.
This model increases rapidly in $z=0-1$, peaks in $z=1-2$, and
gently declines at higher redshifts. A second model remains
approximately constant at $z>2$ and the third model is based on
studies that suggest that the evolving SFR up to $z\approx1$ may
have been overestimated \citep{cow1}, while the rates at high $z$
may have been severely underestimated due to large amounts of dust
extinction \citep{Blain99}.

The model functions of \citet{M2000} and \citet{P2001} have been supplemented by more recent fits to the data (cf. Figure 1). Equation (\ref{sfr}) shows the functional form employed by \citet{hb06} in their fit to the SFR:
\begin{equation}
e(z)  \equiv \frac{\dot\rho_{*}(z)}{\dot\rho_{*}(0)}= \frac{1 + (a_1 z)}{1+ (z/a_2)^{a_3}}
\, \label{sfr}
\end{equation}
where the evolution function $e(z)$ is defined so that $e(0)=1$.

The observed GRB redshift distribution can potentially be used to constrain the independently
derived observation-based SFR density, assuming that the GRB rate directly follows the SFR. Selection effects, such as dust
obscuration in the host galaxies, which affect the observed
SFR, may also modify the observed GRB distribution.
\citet{Blain} show that the ratio of GRBs with and without optical
transients provides an important new probe of the star formation
history of the Universe. Similarly, \citet{Dj03} discuss
the likely association of observed GRB afterglows with
dust obscuration in the host galaxies.  In the following section we review the data and techniques employed to probe the link between the GRB and SFR distributions.

\section{Analysis of the GRB redshift distribution}
\subsection{Pre-{\it Swift} and {\it Swift} redshifts}
From 1997 to 2007 February, about 487 GRBs have been localized using satellite detectors linked to rapid follow-up ground-based optical telescopes, beginning with Beppo-SAX in 1997 and the High Energy Transient Experiment (HETE). The {\it Swift} satellite, a multi-wavelength GRB observatory, has localized about 186 GRBs up to 2006 Nov and a further 35 have been localized by HETE and the satellite detector INTEGRAL. 

From the 487 localizations, 99 redshifts have been obtained with reasonable confidence and 4 of these are associated with short GRBs.  Several `anomalous' GRBs---namely GRBs 980425, 030329, 031203 and 060218---were observed at relatively small redshift (0.0085, 0.168, 0.105, 0.0331 respectively) and were also under-luminous compared to the mean luminosity of long GRBs. Among these four, only the extremely under-luminous GRB 980425 and very long duration GRB 060218 are really peculiar, while the other two may belong to the faint end of the burst distribution. It has also been suggested that GRBs 030329 and 031203 could be accounted for if they are normal bursts seen off-axis, or they may belong to a sub-population of under-energetic GRBs \citep{Coward2005}.

Figure 2 plots the cumulative redshift distribution of 42 and 48 pre-{\it Swift} and {\it Swift} bursts respectively. It is clear that the distributions are biased differently. As of 2006 November, the median redshifts of the two samples are about 1.0 and 2.4 respectively. The discrepancy has been partly explained  by comparing the trigger criteria of the BAT detector onboard {\it Swift} with BATSE. Band (2006) found that the longer temporal accumulation for the trigger of the {\it Swift} BAT makes it more sensitive to long-duration bursts. Also the BAT detects bursts in a lower-energy band than BATSE did, and long-duration bursts are softer, on average, than short-duration bursts. Hence,  the BAT preferentially detects long-duration bursts; because the burst durations are time dilated from cosmic redshift, the BAT is more sensitive to high-$z$ GRBs than was BATSE. Further to this, accurate localizations provided by {\it Swift} and faster response times of telescopes with spectroscopic capabilities have led to more efficient redshift determinations.

\subsection{The GRB luminosity function}
Band (2006) points out that the detectors, BAT and BATSE, select two different subsets of the GRB population. Unifying the two distributions is fraught with difficulty and relies on a complete knowledge of the detector selection function and an accurate model for the GRB luminosity function. Unfortunately the GRB luminosity function is uncertain. Recent studies using the {\it Swift} redshift distribution have attempted to constrain both luminosity function and GRB rate evolution. For example, \citet{Firm04} applied a joint fit to 33 GRB spectroscopic redshifts and 220 redshifts inferred from an empirical variability-luminosity relationship \citep{Fen00} with the aim of constraining both GRB luminosity and the SFR as functions of $z$. They found that the best fit to these data implies that the luminosity is best modelled using a double power-law with slopes $-1.6$ and $-2.6$. They also find that the best fit to the data is achieved using an approximately linear evolution of the GRB luminosity.  In another analysis, \citet{guetta05} derived an empirical double power-law luminosity function assuming parametrized SFR models, but obtained slopes of $-0.1$ and $-2$. 

An example LF is shown below using the double-power law form from \citet{guetta05}. They use lower and upper luminosity limits, $1/\Delta_1$ and $\Delta_2$, respectively, to define the isotropic (not corrected for beaming) luminosity function, $f(L_{\rm iso})$, of GRB peak luminosities $L_{\rm iso}$, in the interval
$\log L_{\rm iso}$ to $\log L_{\rm iso} + d\log L_{\rm iso}$:
\begin{equation}
\label{Lfun}
f(L_{\rm iso})=
\left\{ \begin{array}{ll}
(L_{\rm iso}/L^*)^{\alpha} ,\;\;\;&  L^*/\Delta_1 < L_{\rm iso} < L^* \\
(L_{\rm iso}/L^*)^{\beta}, \;\;\;& L^* < L_{\rm iso} < \Delta_2 L^*\;,
\end{array}
\right. \;
\end{equation}
where $\int f(L_{\rm iso}) dL_{\rm iso}$ is normalized to unity. Following \cite{guetta05}, we employ their parameter values
$\alpha=-0.1$ and $\beta=-2$ for the LF slopes, with $\Delta_1=30$ and $\Delta_2=10$
and an isotropic-equivalent break peak luminosity of $L^*\approx
3.2\times10^{51}$ erg s$^{-1}$. We use this luminosity function in the next section to calculate the GRB scaling function. 

There are many alternative LFs that have been employed to constrain the slope of the LF and the SFR at high $z$. For example Daigne et al. (2006) use a single power law for their Monte Carlo simulations of the GRB redshift distribution, and obtained different slopes assuming 3 different SFR models and GRB peak energy distributions. Assuming the peak GRB energies are described by an Amati relation\footnote{A correlation between the isotropic equivalent GRB energy and the
corresponding peak frequency in the source rest frame---see Amati et al. (2002); Atteia et al. (2003); 
Lamb et al. (2004).} or by a log-normal distribution, they find a power law slope of $-1.5$ to $-1.7$ with minimum and maximum luminosities of $(0.8-3)\times10^{50}$ and $(3-5)\times10^{53}$ erg s$^{-1}$ respectively. They point out that the observed luminosity distribution is biased towards brighter events because the detectors are not sensitive enough to detect the low-luminosity GRBs.

\section{The GRB flux-limited selection function}
The selection function is a fundamental feature of any flux-limited survey with instrumental bias. It incorporates knowledge of the distribution of luminosities of the `target sources' and the sensitivity of the detector. An accurate model of the flux-limited selection function allows one to separate out intrinsic astrophysical parameters, given enough statistics. This is well known in optical and radio surveys of large numbers of galaxies and is equally important in relation to the GRB data obtained from {\it Swift} and other satellites if the data are to be used for cosmological studies. 
 
The fraction of all bursts potentially detectable by a satellite depends on many factors, including detector area, detector efficiency, angular field of view and instrumental noise. In the absence of instrumental selection effects, such as the temporal sensitivity to GRBs (e.g., Band 2006), the detection fraction depends completely on the flux sensitivity threshold of the instrument--see e.g., \citet{Schaefer1,
Schmidt01b,Norris02}. 
\subsection{The Band function}
The GRB photon spectral luminosity can be approximated by the Band et al. (1993) parametrized fit, expressed as:
\begin{eqnarray}
S(E)=A\times  
\left\{ \begin{array}{ll}
%\begin{cases}
\displaystyle{
\left(\frac{E}{100\,{\rm keV}}\right)^\alpha\exp\left[\frac{
(\beta-\alpha)E}{E_b}\right]},\;{E<E_b} \\
\displaystyle{
\left(\frac{E_b}{100\,{\rm keV}}\right)^{\alpha-\beta} 

\left(\frac{E}{100 \,{\rm keV}}\right)^\beta}\exp{(\beta-\alpha)}, \;{E\geq E_b}\;,
%\end{cases}
\end{array} \right.
\end{eqnarray}
with $S(E)$ the source rest-frame photon spectral (or differential) luminosity, so that $A$ is in photons s$^{-1}$ keV$^{-1}$; $S(E)dE$ is the number of photons emitted per second in the energy interval $E$ to $E+dE$. Integrating $ES(E)dE$ over some energy band then gives the total (all angles) photon bolometric luminosity in that band. Assuming isotropic emission, dividing by the surface area of a sphere centered on the source converts this to the photon flux in that band, with units photons s$^{-1}$ cm$^{-2}$. The energy spectral indices, $\alpha$ and $\beta$ can be obtained from the BATSE spectral catalogue of \cite{preece} and have typical values $-1$ and $-2.25$ respectively, with a spectral break at $E_b=511$ keV \citep{nat05} but we note that any value for $E_b \sim 500$ keV is a reasonable approximation.

\subsection{Flux limited GRB detection}
As shown by  \citet{guetta05}, the flux limit, $P_{\rm lim}(L_{\rm lim},z)$, for a GRB of peak luminosity $L_{\rm lim}(z)$ and photon spectral luminosity $S(E)$ at a bolometric luminosity distance $d_{\mathrm L}(z)$, is
\begin{equation} \label{flux}
P_{\rm lim}(L_{\rm lim},z) = \frac {L_{\rm lim}} {4 \pi d^2_{\mathrm L}(z)} \frac{\int_{(1+z) E_1 }^{(1+z) E_2 }S(E) dE}
 {\int_{E_1}^{E_2}S(E) dE}\;,
\end{equation}
where $E_1<E<E_2$ defines the detected energy band and the $(1+z)$ factor in the integration limits accounts for the redshift of the rest frame source spectrum.
Typical values for the flux sensitivity limit, $P_{\rm lim}$, of the detectors BATSE and {\it Swift} are 10$^{-7}$ and 10$^{-8}$ ergs cm$^{-2}$ s$^{-1}$ respectively with corresponding sensitivity bands $50-300$ and $15-150$ keV.

The luminosity distance used above is related to the proper distance $d_\mathrm{P}(z)$ by  
\begin{equation}
d_{\mathrm L}(z)=(1+z)d_\mathrm{P}(z)\;,
\end{equation}
and the proper distance is defined by
\begin{equation}
H_0 d_\mathrm{P}(z)/c =\int_0^{z}
{\mathrm{d}x\hspace{0.05cm}}\hspace{-0.02cm}/h(x,\Omega_{\mathrm{M}})\;.
\end{equation}
where $H_0$ is Hubble's constant and the dimensionless Hubble parameter $h(z,\Omega_{\mathrm{M}})$ is given by
\begin{equation}
h(z,\Omega_{\mathrm{M}})=\sqrt{1-\Omega_{\mathrm{M}}+(1+z)^3\Omega_{\mathrm{M}}}
\end{equation}
for a spatially flat cosmology.

The fraction of detectable GRBs, or scaling function (those observed with peak flux  $>P_{\rm lim}$), is
\begin{equation}\label{grbrate}
\psi_{\rm GRB}(z)= \int_{L_{\rm lim}(P_{\rm lim},z)}^{\infty}f(L_{\rm iso}) {\rm d}L_{\rm iso} \;,
\end{equation}
where $L_{\rm lim}(P_{\rm lim},z)$ is obtained by solving equations (\ref{flux}). The GRB scaling function, Eq. (\ref{grbrate}), is the inverse of the detector selection function. Figure 3 plots $\psi_{\rm GRB}(z)$ using the LF defined by equation (\ref{Lfun}) with 
slopes $\alpha=-0.1$ and $\beta=-2$ and using the {\it Swift} detector flux limit 10$^{-8}$ ergs cm$^{-2}$ s$^{-1}$. The curve highlights how GRB detection evolves strongly with redshift.

The GRB scaling function is fundamental to any model used to fit to GRB redshift data. It encompasses both the GRB luminosity function uncertainty and the detector response to a spatial distribution of transient fluxes. But it does not include other selection biases from either the satellite detectors or the redshift measurements from ground-based telescopes. This issue will be discussed further in Section \ref{complete}.

\subsection{The GRB probability distribution function}
There are a variety of approaches in modeling and simulating the GRB redshift distribution. A straightforward method that highlights the key parameters is described below. 
The all-sky event rate distribution of transient cosmological
events, such as core-collapse supernovae and GRBs,
can be expressed by the event-rate equation (e.g. Coward, Burman \& Blair 2001):
%\citep{Coward2001}:
\begin{equation}
dR/dz = 4\pi (c^3r{_0}/H_0{^3})e(z)F(z)/(1+z)\label{EQdR}
\end{equation}
where $R(z)$ is the all-sky event rate, as observed in our local
frame, for sources out to redshift $z$. The factor
$(c^3r{_0}/H{_0}^3)$ has the dimensions of inverse time, as does
the event rate $R$; the factor $r{_0}$ is the ``local" rate
density of events, the event rate per unit volume at the present
epoch. The $(1 + z)$ denominator in (\ref{EQdR}) accounts for the
time dilation of the observed rate by cosmic expansion, converting
a source-count equation to a rate equation.  Equation (\ref{EQdR}) represents the event-rate in shells of
redshift $dz$; its form is defined by the expression
$e(z)F(z)/(1+z)$. 

The function $F(z)$ is determined by the cosmological model. For
flat-$\Lambda$ cosmologies (Peebles 1993, p. 332):
\begin{equation}
 F(z) = [\,H_0 d_\mathrm{P} (z)/c\,]^{\,2}/h(z,\Omega_{\mathrm{M}})\;.
\label{EQ Fcos}
\end{equation}

We assume a flat-$\Lambda$
cosmology ($\Omega_{\mathrm M}=0.3$, $\Omega_{\mathrm
\Lambda}=0.7$) with $H_0=70$ km s$^{-1}$ Mpc$^{-1}$.

The probability distribution function for the potentially
observable GRB rate variation with $z$ can be expressed as
\begin{equation}\label{pdf1}
p(z) = {\rm N}^{-1}\psi_{\rm GRB}(z)e(z)F(z)/(1+z)\;,
\end {equation}
with normalizing factor
\begin{equation}
{\rm N} \equiv \int_{0}^{z_{\rm
max}}\psi_{\rm GRB}(z)[e(z)F(z)/(1+z)]{\rm d}z \;.
\end{equation}

Figure \ref{fig3} plots equation (\ref{pdf1}) using the same scaling functions as shown in figure \ref{lumplot}. For definiteness we assume GRB rate evolution traces a SFR model that peaks in $z=1-2$, and remains approximately constant at $z>2$. Equation (\ref{pdf1}) with the normalizing factor $N^{-1}$ replaced by $4\pi (c^3r{_0}/H_0{^3}) \epsilon$ is the differential GRB rate as observable by a satellite with a flux limited scaling function $\psi_{\rm GRB}$; the $\epsilon$ factor defines the field of view and efficiency of the detector and $e(z)$ is the evolving GRB rate density normalized to unity at $z=0$.

\subsection{Utilizing the GRB redshift distribution}
The fact that GRB redshifts have been measured from the local Universe to $z=6.9$ provides strong motivation for using the redshift distribution to constrain the evolution of the GRB rate density. Previous studies have focussed on constraining GRB rate evolution by assuming that they trace the evolving SFR. \citet{Firm04} find that their best fit for the SFR, from a joint fit using 33 GRB spectroscopic redshifts and 220 redshifts inferred from an empirical variability- luminosity law, implies a general shape that is consistent with other independent tracers of the SFR, such as the rest-frame luminosity densities of galaxies. 

The study by Daigne et al. (2006) employed the log $N$ -- log$P$ and peak energy distribution of BATSE GRBs and the HETE2 fraction of X-ray rich flashes to constrain GRB evolution. Instead of fitting directly to these distributions, Daigne et al. used Monte Carlo simulations to study the effect of the uncertainties discussed above on the GRB rate. They find a GRB to supernova rate ratio of 1:$10^5-10^6$, assuming no beaming factor, but do not attempt to quantify the uncertainty in this range.
The most significant result of their analysis is the prediction of a positive GRB evolution that exceeds the best fit models for the evolving SFR. They quote a GRB rate density at $z=7$ of 6--7 times larger than at $z=2$. Daigne et al. interpret this as evidence for GRBs not tracing the evolving SFR but rather tracing the evolution of GRB progenitor rate efficiency. This is an attractive idea given that several studies indicate that some GRBs are associated with dwarf irregular low-metallicity galaxies \citep{fruch06}. 

Another study, by \cite{le06}, uses a GRB model with a non-evolving luminosity function to fit to the redshift and jet opening angle distributions to constrain GRB evolution. The authors use seven variable parameters in their data fits, with four of them partly constrained by observation. Their best fit to a luminosity function, assuming a single power law, finds a slope $-3.25$, an intrinsic beaming factor of $34-42$ and a GRB rate of one in every 600,000 years in a Milky Way type galaxy. Similarly to Daigne et al., they find positive GRB evolution at high $z$, but their calculated GRB to supernovae ratio of about $10^4$ is $1-2$ orders of magnitude smaller.  Truong \& Dermer point out that they could only find statistically acceptable joint fits to the data if the GRB rate density is increasing monotonically at $z>5$, but they do not attempt to provide confidence limits to the fits.

\section{Open issues}
\subsection{An evolving GRB rate density?} 
Daigne et al. (2006) and \cite{le06} reach similar conclusions for the evolution of the GRB rate density; they find that the observed GRB rate at high $z$ is higher than that expected.  \cite{Firm04} reach a similar conclusion, but explain this higher than expected rate as evidence for an evolving GRB luminosity function. These results are very interesting and may be important to our understanding of how GRBs formed during earlier cosmological epochs. If they are robust, there are several possible reasons why the GRB rate is relatively excessive at high-$z$. The number density of GRB progenitors may have been enhanced in earlier cosmological epochs because of the expected higher density of massive stars in low-metallicity host environments.

Very recent work has shown that long GRBs are more probable in low-metallicity galaxies: \citet{fruch06} used Hubble Space Telescope images of 42 GRB afterglows and their host galaxies in $z<1.2$, and found that dwarf irregular galaxies are the predominant host galaxy type. In addition they found that the GRBs in their sample are located in the brightest parts of their hosts, not where the blue light characteristic of active star formation is most intense. \citet{stan06} have shown that host galaxies of the five long GRBs with redshifts $z\le0.25$ have distinctly low oxygen abundance in comparison with star-forming galaxies in $0.005<z<0.2$ from the Sloan Digital Sky Survey.

Dwarf irregular galaxies undergo prolonged but gradual star formation, and are predicted to contain less than 10\% of their metallicity in stars \citep{Cal1}. For dwarf irregulars, the stellar metal density is a slowly increasing function throughout their histories, implying a continuous but relatively low SFR. If GRBs are more probable in low-metal enriched environments, GRB evolution should track the evolving low-metallicity component of the SFR. 

There is also the strong possibility that the GRB population consists of several intrinsically different populations. The observation of the so-called `anomalous' under-luminous GRBs, namely GRBs 980425, 030329, 031203 and 060218, may imply a more numerous population based on their relatively small redshifts. Does this imply a different class of progenitor from that of the classical GRBs? What are the implications for the low-energy tail of the luminosity function?

\subsection{The complete GRB selection function}\label{complete}
Knowledge of the GRB selection function, $\psi_{\rm GRB}(z)$, described by Equation (\ref{grbrate}), is  defined by a GRB LF, a flux-limited sample and a choice of cosmology dependent luminosity distance. It would represent the `complete selection function' in the absence of all other instrumental selection biases, which unfortunately is definitely not the case. There is clear evidence that instrumental biases from the satellite observations and the redshift measurements result in a marked distortion of the `true' GRB redshift distribution.

Only about 50\% of {\it Swift} GRBs have been observed with an optical afterglow,  despite the rapid localization and follow up by telescopes capable of spectroscopy. This leaves a significant fraction of the observed population with no possibility of observing a redshift directly from the afterglow. One possible reason for this deficiency is dust extinction of the afterglow in the GRB host galaxy, but this may not account for all of the deficiencies. \cite{Fi06} showed that the different redshift distributions obtained using BeppoSax/HETE2 and {\it Swift} were at least partially due to the sensitivity of the different instruments, localisation efficiency and efficiency of obtaining a spectroscopic redshift of the afterglow or of the host galaxy. The afterglow brightness is an important factor in obtaining a spectroscopic redshift.
The redshifts that have been measured come from bright GRBs, implying the measured GRB redshift distribution is a sub-population of the bright part of the GRB LF. 

An important question is how selection biases affect the fitting procedures employed to extract the SFR and/or GRB rate evolution using redshift data? If this bias is not accounted for in the GRB scaling function, the model for the observed redshift distribution will be biased.  The \cite{Band06} analysis showed that {\it Swift} has increased sensitivity to high redshift bursts because of their longer duration. This selection effect may manifest in the redshift distribution as an evolution of the GRB rate as a function of redshift. The inability to account for these biases in the model fits to the redshift distribution could easily result in misinterpreting fitting results. In addition, the high-$z$ regime where this may be a problem is also where there is the most uncertainty in the GRB rate density and SFR. 

To account for the most significant biases requires a `complete GRB selection function'. An expression for such a function could consist of a product of normalized scaling functions for the main classes of biases\footnote{We note the complete scaling function, $\psi_{\rm complete}(z)$, or its components, $\psi_{\rm GRB}(z)$ and $\psi_{\rm satellite}(z)$, are sometimes termed `detection efficiency' curves.}: 
\begin{equation}
\psi_{\rm complete}(z)=\psi_{\rm GRB}(z)\psi_{\rm satellite}(z)\psi_{\rm redshift}(z)\;,
\label{select1}
\end{equation}
where $\psi_{\rm satellite}(z)$ would account for the bias discussed by Band (2006) and $\psi_{\rm redshift}(z)$ would incorporate a model for the `redshift dropout'  from the difficulty of obtaining resolved lines in different redshift regimes (Firore et al. 2006). Various attempts have been made to at least minimize these biases, including selecting only those redshifts that pass certain criteria, such as the source position in relation to Galactic dust obscuration.

The use of Monte Carlo simulations to study the propagation of uncertainty from the GRB selection function to the fitting parameters for GRB evolution will provide a better understanding of the uncertainty in the best-fit models. This has already been performed using different LFs, but has not been fully extended to instrumental biases, partly because of the complexity of the biases. Nonetheless, unless the complete GRB selection function can be constrained to a better precision, it is unlikely that it will be possible to discriminate between different GRB evolutionary scenarios based on the redshift distribution alone.

\section{Concluding remarks}
Presently there is no `complete' GRB selection function to account for the biases emerging in the GRB redshift data. Nonetheless there is an urgent need to quantify selection effects that plague the observed GRB redshift distribution.  The result of this uncertainty in the redshift distribution is that it is difficult to differentiate between an evolving GRB luminosity function, enhanced rate density and the shape of the SFR in $z>2-3$. It is very difficult to place quantitative confidence limits on the fitted GRB evolution parameters using the redshift distribution because the uncertainties inherent in the models are not fully quantified, especially $\psi_{\rm redshift}(z)$. Previous studies exploit both the brightness and redshift distributions to explore different GRB evolutionary scenarios and estimate parameter limits based on a range of plausible models. If all the instrumental biases from the redshift distribution are accounted for, we can explore the physical GRB rate evolution parameter space with confidence and our main limitation will be small number statistics, a problem that will inevitably be solved by acquiring more data from {\it Swift} and other satellites.

\section{Acknowledgements}
D. M. Coward is supported by the Australian Research Council. D. M. Coward thanks participants of the workshop `Recent Developments in the  Study of Gamma Ray Bursts'--Royal Society, London, September 2006, for helping clarify the importance of biases inherent in the GRB redshift distribution. He also thanks Dr Ron Burman and Prof. David Bair for constructive feedback and Prof. E.P.J. van den Heuvel  for informative discussions during his visit in November 2006 to the University of Western Australia. Comments from the referee have been invaluable and helped the author clarify several key issues.

\newpage
\begin{figure}
\includegraphics[scale=0.5,angle=-90]{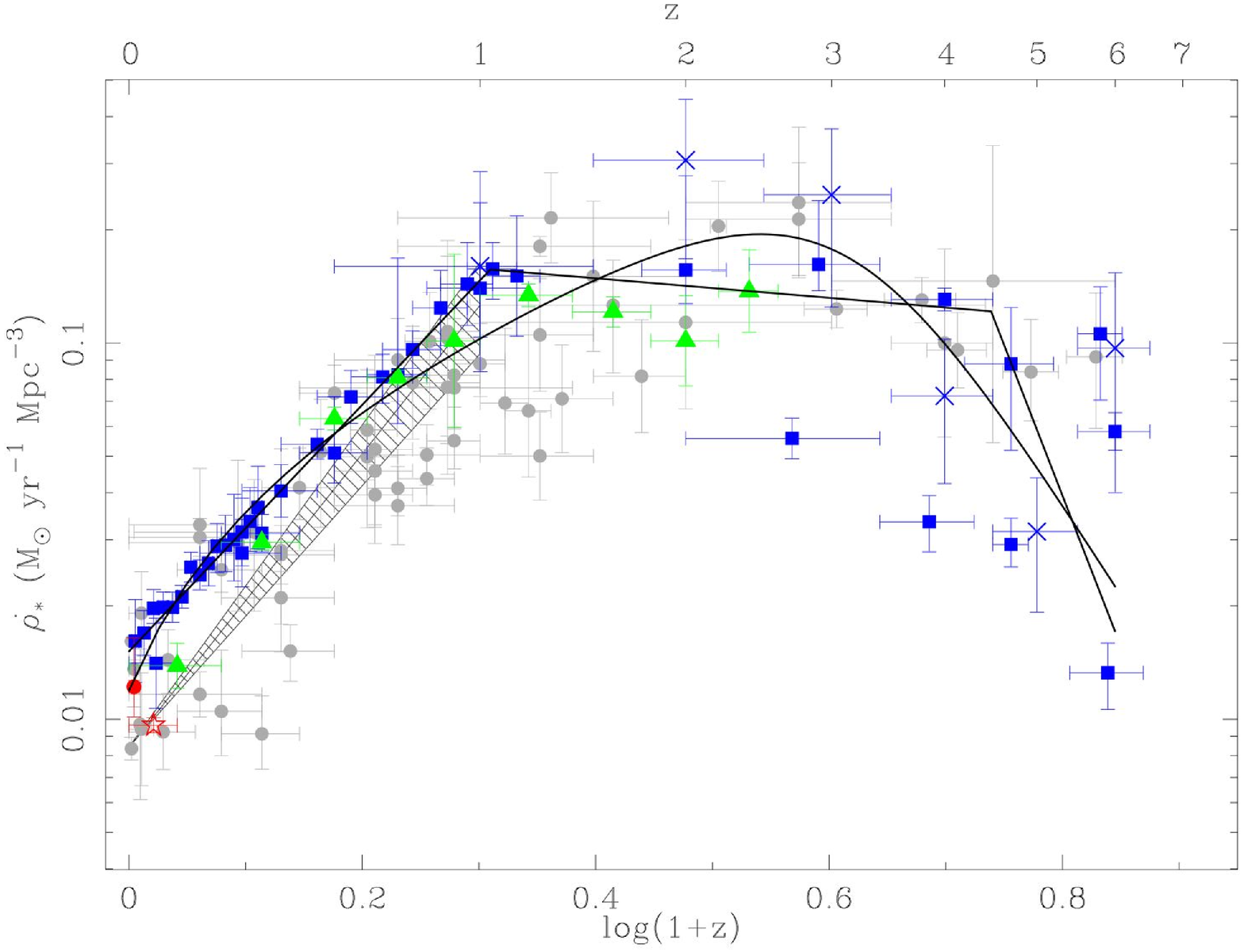} \caption{Evolution of the SFR density with redshift reproduced from \cite{hb06}. They have compiled data from various sources (see references therein) including  FIR ($24\,\mu$m)---hatched region \& triangles, UV---squares, radio(1.4 GHz)---open red star and UDF estimates---crosses. The solid lines are their fits to these data using parametrized fitting functions.
 }\label{sfr}
\end{figure}

\begin{figure}
\includegraphics[scale=1]{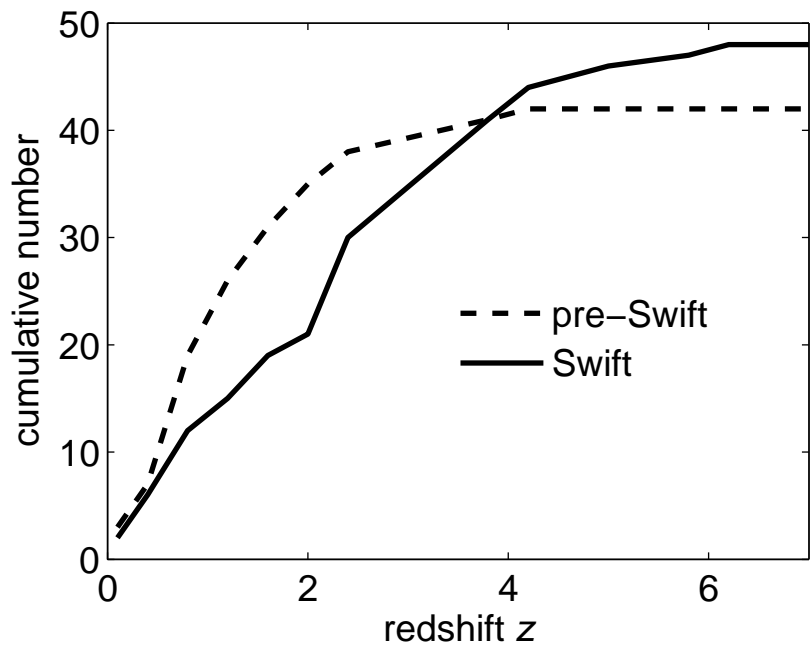} \caption{Plots of the cumulative redshift distributions using 42 and 48 pre-{\it Swift} and {\it Swift} bursts respectively. The median redshift of the two distributions are 1.0 and 2.4 respectively, indicating the presence of different biasses.  }\label{fig2}
\end{figure}

\begin{figure}
\includegraphics[scale=1.2]{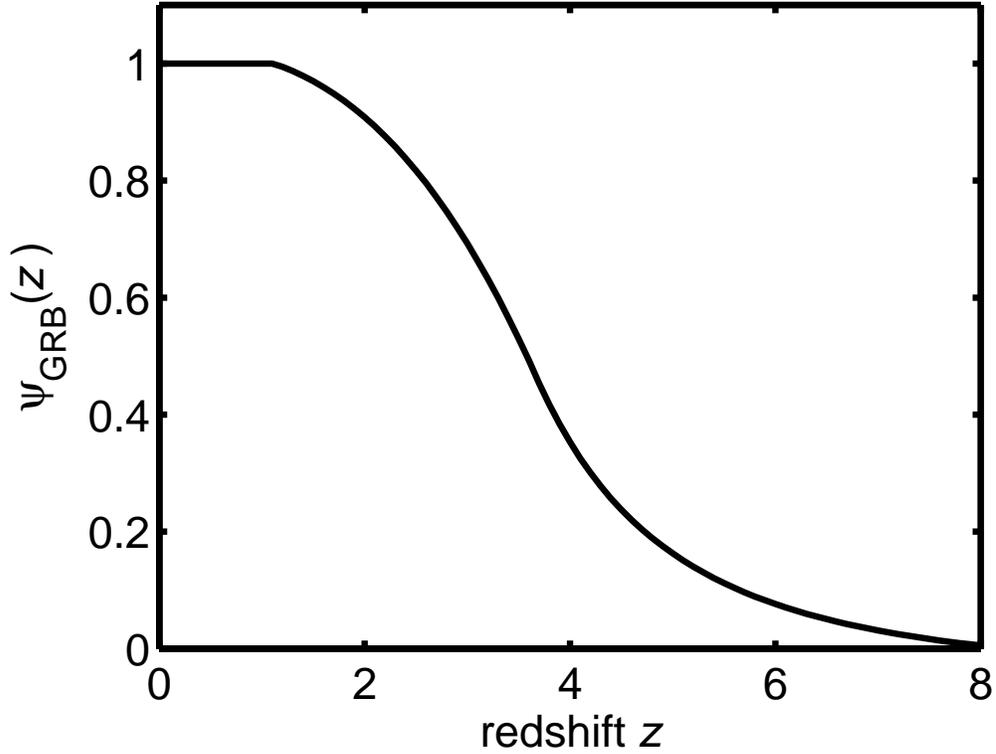} \caption{A plot of the GRB scaling function, $\psi_{\rm GRB}(z)$, using the LF expressed in equation (\ref{Lfun}) and the peak observable flux from equation (\ref{flux}) assuming a limiting detector sensitivity similar to {\it Swift} of 10$^{-8}$ cm$^{-2}$ s$^{-1}$. The Band function with spectral slopes $\alpha=-1$ and $\beta=-2.25$, integrated over the 50--300 keV band, is used to define the source luminosity. } \label{lumplot}
\end{figure}

\begin{figure}
\includegraphics[scale=1]{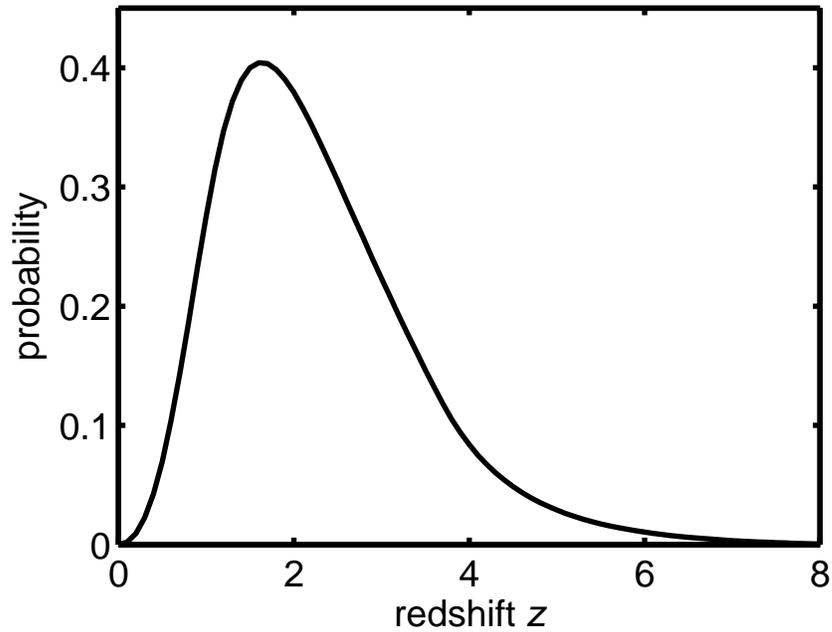} \caption{Plot of the probability distribution of GRBs as a function of redshift, assuming the scaling function plotted in figure \ref{lumplot}. GRB rate evolution is assumed to follow the SFR and for definiteness we use a SFR model from \cite{P2001} that peaks in $z=1-2$, and remains approximately constant at $z>2$. The model curves do not account for selection biases associated with the satellite detectors or the redshift determinations.  }\label{fig3}
\end{figure}

%\appendix
%\section{APPENDIX}

\end{document}